\documentclass[pdflatex,sn-mathphys-num]{sn-jnl}

\usepackage{graphicx}%
\usepackage{multirow}%
\usepackage{amsmath,amssymb,amsfonts}%
\usepackage{amsthm}%
\usepackage{mathrsfs}%
\usepackage[title]{appendix}%
\usepackage{xcolor}%
\usepackage{textcomp}%
\usepackage{manyfoot}%
\usepackage{booktabs}%
\usepackage{algorithm}%
\usepackage{algorithmicx}%
\usepackage{algpseudocode}%
\usepackage{listings}%

\usepackage{physics}
\usepackage{subcaption} 

\theoremstyle{thmstyleone}%
\theoremstyle{thmstyletwo}%

\theoremstyle{thmstylethree}%

\begin{document}

\title{Pair distribution function of the cell fluid model with Curie-Weiss interaction}


\author*{\fnm{R.~V.} \sur{Romanik}}\email{romanik@icmp.lviv.ua}

\author{\fnm{O.~A.} \sur{Dobush}}\email{dobush@icmp.lviv.ua}

\author{\fnm{M.~P.} \sur{Kozlovskii}}\email{mpk@icmp.lviv.ua}

\author{\fnm{I.~V.} \sur{Pylyuk}}\email{piv@icmp.lviv.ua}

\author{\fnm{M.~A.} \sur{Shpot}}\email{shpot.mykola@gmail.com}
\affil{\orgname{Yukhnovskii Institute for Condensed Matter Physics of the National Academy of Sciences of Ukraine}, \orgaddress{
		\city{Lviv}, \postcode{79011}, \country{Ukraine}}}
	

\abstract{In this paper we present results for the pair distribution function for the cell fluid model with Curie-Weiss interaction. As a supplementary result, one- and two-particle densities are calculated.}

\keywords{Cell fluid model, Curie-Weiss interaction, Pair distribution function, $n-$particle density}



\maketitle

\section{Introduction}
Various models of many-particle systems interacting via soft potentials have  been studied in the literature. Examples include the penetrable spheres model~\cite{MW89,LWL98,Schmidt99,FLL00,Likos01}, the Gaussian core model~\cite{Stillinger76,LLWL00,LBH00,LBHM00,BLHM01}, and the generalized exponential model of index 4 (GEM-4)~\cite{Prestipino14,PGT15}. When treating such systems within lattice-gas frameworks, it is natural to consider lattice gases with multiple cell occupancies~\cite{FHL04,FL18}. 

In this paper, we study the cell fluid model introduced in~\cite{KKD18,KKD20}, which represents a multiple-occupancy model, and derive an exact expression for its pair distribution function in the case of Curie-Weiss interaction. In the course of analysis, we also obtain the one- and two-particle densities. The results confirm that the cell fluid model with Curie-Weiss interaction is essentially a mean-field fluid.

\section{\label{sec:model} Model}
An open system of point particles is considered in volume $V\subset\mathbb R^3$ in three space dimensions. The total volume $V$ is partitioned into $N_v$ non-overlapping congruent cubic cells $\Delta_l$, $l\in\{1,...,N_v\}$, each of volume $v$, such that $V$ is the union of all $\Delta_l$'s:
\begin{eqnarray}\label{volume}
	V = \bigcup_{l=1}^{N_v}\Delta_l,\qquad
	\Delta_l \cap \Delta_{l'} = \emptyset, \text{ if } l \neq l',
	\\ \qquad\mbox{and}\quad
	V = v N_v.
\end{eqnarray}
The interaction energy between particles in configuration $\gamma_N = \{{\vb r}_1, ..., {\vb r}_N\}$, where ${\vb r}_i$ is the space coordinate of the $i$-th particle and $N$ is the number of particles in the configuration $\gamma_N$, is defined as follows
\begin{equation*}
	W_{N_v}(\gamma_N) = \frac{1}{2} \sum_{{\vb r}_i,{\vb r}_j \in \gamma} \Phi_{N_v} ({\vb r}_i, {\vb r}_j)
\end{equation*}
where $\Phi_{N_v}$ is given by the Curie-Weiss interaction
\begin{equation}
	\label{def:curie-weiss-pot}
	\Phi_{N_v}(\vb{r}_i, \vb{r}_j) = -\frac{J_1}{N_v} + J_2\sum_{l=1}^{N_v} \mathbb{I}_{\Delta_l}(\vb{r}_i) \mathbb{I}_{\Delta_l}(\vb{r}_j).
\end{equation}
The first term in $\Phi_{N_v}$ describes a global Curie-Weiss (mean-field-like) attraction for any pair of particles in the system.
The strength of this attraction is controlled by an energy parameter $J_1 > 0$. The second term in $\Phi_{N_v}$ describes a local repulsion between two particles contained within the same cell $\Delta_l$ and is characterized by the parameter $J_2 > 0.$
Here, $\mathbb{I}_{\Delta_l}(\vb{r})$ is the indicator function of a cell $\Delta_l$,
\begin{equation}
	\label{def:I}
	\mathbb{I}_{\Delta_l} (\vb{r}) = \left\{
	\begin{array}{ll}
		1, \quad \vb{r} \in \Delta_l,
		\\
		0, \quad \vb{r} \notin \Delta_l.
	\end{array}
	\right.
\end{equation}
We refer the reader to Refs.~\cite{KKD18,KKD20,RDKPS25arxiv} for a more rigorous definition of the model. The phase diagrams in both the temperature-density and pressure-temperature planes are presented in~\cite{KD22}.

Let us briefly introduce notation to be used in this paper, and summarize the known results for the model.

The grand partition function is expressed as follows
\begin{equation*}
	\label{gpf}
	\Xi = \sum_{N=0}^{\infty}\frac{\zeta^N}{N!} \int_{V} \dotsc \int_{V} \exp\left[-\beta W_{N_v}(\gamma_N)\right] {\rm d} {\vb r}_1 \dotsc {\rm d} {\vb r}_N
\end{equation*}
where $\zeta$ is the activity
\begin{equation*}
	\zeta = \frac{\exp(\beta \mu)}{\Lambda^3},
\end{equation*}
$\beta = k_{\rm B} T$ the inverse temperature, $k_{\rm B}$ the Boltzmann constant, $T$ the temperature, $\mu$ the chemical potential, $\Lambda = (2\pi\beta\hbar^2/m)^{1/2}$ the de Broglie thermal wavelength, $\hbar$ the reduced Planck constant, $m$ the mass of a particle. In the grand partition function~\eqref{gpf} the integration goes over all configurations with $N$ particles and then the summation goes over all positive integer values of $N$.

We use the following standard dimensionless variables: the reduced temperature $T^*=k_{\mathrm{B}}T/J_1$, the reduced pressure $P^* = P v/J_1$, the reduced chemical potential $\mu^* = \mu/J_1$, the reduced particle density $\rho^* = \frac{\langle N \rangle}{V} v = \frac{\langle N \rangle}{N_v}$, with $P$ being the pressure, and $\langle N \rangle$ the average number of particles in the system, where the brackets denote a grand-canonical ensemble average. The particle density $\rho = \frac{\langle N \rangle}{V}$ is also used throughout the paper.

For stability of interaction the following condition must hold
$J_2 > J_1.$
For this reason we introduce notation
$a = J_2/J_1.$ and set $a=1.2$ in the current work.

The exact asymptotic expressions for the grand partition function were obtained in~\cite{KKD18,KKD20}. Here, we present the result in the form presented in~\cite{RDKPS25arxiv}:
\begin{equation}
	\Xi \simeq c_{N_v} \exp[N_v E(T^*, \mu^*; \bar{y}_{\rm max})],
\end{equation}
which is asymptotically exact for the limit of $N_v \to \infty$. Here $\bar{y}_{\rm max}$ is a function of both $T^*$ and $\mu^*$ and is found from the condition of global maximum for quantity $E$, which, in turn, is defined as follows:
\begin{equation*}
	E(T^*, \mu^*; y) = -\frac{y^2}{2}T^* + \ln K_0(T^*,\mu^*; y),
\end{equation*}
where $K_0(T^*,\mu^*; y)$ is the $0$-th member of the following family of special functions
\begin{eqnarray}\label{def:Kj}
	K_j(T^*,\mu^*;y) = \sum_{n=0}^{\infty} \frac{n^j \left(v^* T^{*3/2} \right)^n}{n!} \exp[\left(y+\frac{\mu^*}{T^*}\right)n - \frac{a}{2T^*}n^2]
\end{eqnarray}
for $j=0,1,2,\ldots$, and $v^* = v/\lambda^3$, $\lambda = (2\pi\hbar^2/mJ_1)^{1/2}$. Quantity $c_{N_v}$ does not play any important role in the thermodynamic limit, therefore we do not provide its explicit expression here.

For detailed investigation of the conditions necessary for the global maximum of $E$ and the derivation of relations for $\bar{y}_{\rm max}$, we direct the interested readers to works~\cite{KKD18,KKD20,KD22,RDKPS25arxiv}. In this paper, we employ the methodology developed and the results derived in those works to study structure properties of the model, in particular one- and two-particle densities, and the pair distribution function.

\section{\label{sec:init-pair-distribution} Pair distribution function}
The $n$-particle distribution function is defined via the $n$-particle densities as follows (for reference, see Section~2.6 in~\cite{HansenMcDonald13})
\begin{equation*}
	\label{def:g_n}
	g^{(n)}({\vb r}_1, \dotsc, {\vb r}_n)=\frac{\rho^{(n)}({\vb r}_1, \dotsc, {\vb r}_n)}{\prod_{i=1}^{n}\rho^{(1)}({\vb r}_i)},
\end{equation*}
where the $n$-particle density defined in the grand canonical ensemble is
\begin{eqnarray*}
	\label{def:rho_n}
	\rho^{(n)}({\vb r}_1, \dotsc ,{\vb r}_n) = \frac{1}{\Xi}\sum_{N=n}^{\infty} \frac{\zeta^N}{(N-n)!} \int\exp\left[-\beta W_{N_v}(\gamma_N)\right] {\rm d} {\vb r}_{n+1} \dotsc {\rm d} {\vb r}_N.
\end{eqnarray*}
In this work, our goal is to calculate the pair distribution function $g^{(2)}({\vb r}_1, {\vb r}_2)$
\begin{equation*}
	g^{(2)}({\vb r}_1, {\vb r}_2) = \frac{\rho^{(2)}({\vb r}_1, {\vb r}_2)}{\rho^{(1)}({\vb r}_1) \rho^{(1)}({\vb r}_2)}.
\end{equation*}
Therefore, we first need to calculate the one- and two-particle densities.

\subsection{One-particle density}

Let us first calculate $\rho^{(1)}$. By definition
\begin{equation}
	\label{def:rho1}
	\rho^{(1)} ({\vb r}_1) = \Xi^{-1} \sum_{N=1}^{\infty} \frac{\zeta^N}{(N-1)!} \int \exp\left[-\beta W_{N_v}({\vb r}_1, \dotsc, {\vb r}_N)\right] {\rm d} {\vb r}_2 \dotsc {\rm d} {\vb r}_N.
\end{equation}
Applying the method described in~\cite{KKD20}, we obtain the following result
\begin{equation*}
	\label{eq:rho1}
	\rho^{(1)} ({\vb r}_1) = \frac{K_1(T^*, \mu^*; \bar{y}_{\rm max})}{v K_0(T^*,\mu^*; \bar{y}_{\rm max})} = \frac{\rho^*(T^*,\mu^*)}{v}
\end{equation*}
where we used the following equality~\cite[Eq.~(42)]{RDKPS25arxiv}
\begin{equation}
	\label{eq:rho_in_y}
	\rho^* = \frac{K_1(T^*, \mu^*; \bar{y}_{\rm max})}{K_0(T^*,\mu^*; \bar{y}_{\rm max})}.
\end{equation}
Therefore, for the considered cell fluid model with Curie-Weiss interaction we obtained the result known for homogeneous systems: the single-particle density is equal to the average particle density, see~\cite[Eq.~(2.6.5)]{HansenMcDonald13}:
\begin{equation}
	\rho^{(1)}({\vb r}_1) = \rho.
\end{equation}

In the present theory, the average value of the cell-occupation number $n$ can be calculated using the probability measure $Q_{T^*,\mu^*}(n)$~\cite{KKD20}:
\begin{equation*}
	Q_{T^*,\mu^*}(n) = \frac{1}{K_0(T^*, \mu^*; \bar{y}_{\rm max}) n!} \left(v^* T^{*3/2}\right)^n \exp\left[\left(\bar{y}_{\rm max} + \frac{\mu^*}{T^*} \right)n - \frac{a}{2T^*}n^2 \right],
	\quad n \in \mathbb{N}_0,
\end{equation*}
so that
\begin{equation*}
	\langle n \rangle_{Q} = \sum_{n=0}^{\infty} n Q_{T^*, \mu^*}(n),
\end{equation*}
where $\langle (\ldots) \rangle_Q$ denotes averaging with respect to the probability measure $Q_{T^*, \mu^*}(n)$.
Hence, we can write the following equalities:
\begin{equation}
	v \rho^{(1)}({\vb r}_1) = \langle n \rangle_{Q} = \rho^*.
\end{equation}

\subsection{Two-particle density}

Let us now calculate the two-particle density, defined as
\begin{equation*}
	\rho^{(2)} ({\vb r}_1, {\vb r}_2) = \Xi^{-1} \sum_{N=2}^{\infty} \frac{\zeta^N}{(N-2)!} \int \exp\left[-\beta W_{N_v}({\vb r}_1, {\vb r}_2, \dotsc, {\vb r}_N)\right] {\rm d} {\vb r}_3 \dotsc {\rm d} {\vb r}_N.
\end{equation*}
To deal with this expression, we again follow the methodology developed in~\cite{KKD20}. However, it is important to note that the result for $\rho^{(2)}({\vb r}_1, {\vb r}_2)$ varies depending on whether ${\vb r}_1$ and ${\vb r}_2$ belong to the same cell or to different cells.

When ${\vb r}_1$ and ${\vb r}_2$ belong to different cells, the result for $\rho^{(2)}({\vb r}_1, {\vb r}_2)$ is as follows
\begin{eqnarray}
	\rho^{(2)} ({\vb r}_1, {\vb r}_2) & = & \left[\frac{K_1(T^*, \mu^*; \bar{y}_{\rm max})}{v K_0(T^*, \mu^*; \bar{y}_{\rm max})}\right]^2
	\nonumber \\
	& = & \frac{\rho^{*2}}{v^2} = \frac{\langle n \rangle_{Q}^2}{v^2}.
\end{eqnarray}
From this expression and Eq.~\eqref{eq:rho1} the result for $g^{(2)}({\vb r}_1, {\vb r}_2)$ immediately follows
\begin{equation}
	\label{g2_diff}
	g^{(2)}({\vb r}_1, {\vb r}_2) = 1, \quad \text{if } \nexists \Delta_l ({\vb r}_1, {\vb r}_2 \in \Delta_l),
\end{equation}
that is, when ${\vb r}_1$ and ${\vb r}_2$ belong to different cells, the pair distribution function $g^{(2)}({\vb r}_1, {\vb r}_2)$ always equals $1$, i.e. is equal to the ideal-gas value.

When ${\vb r}_1$ and ${\vb r}_2$ belong to the same cell, the result for $\rho^{(2)}({\vb r}_1, {\vb r}_2)$ is as follows (cf. from~\cite[Eq.~(2.6.4)]{HansenMcDonald13})
\begin{equation}
	\rho^{(2)}({\vb r}_1, {\vb r}_2) = \frac{1}{v^2} \sum_{n=0}^{\infty} n(n-1) Q_{T^*, \mu^*}(n) 
	= \frac{\langle n(n-1) \rangle_{Q}}{v^2}.
\end{equation}
Therefore, for $g^{(2)}({\vb r}_1, {\vb r}_2)$ one gets
\begin{equation}
	\label{g2_same}
	g^{(2)}({\vb r}_1, {\vb r}_2) = \frac{\langle n(n-1) \rangle_{Q}}{\langle n \rangle_{Q}^2}, \quad \text{if } \exists \Delta_l ({\vb r}_1, {\vb r}_2 \in \Delta_l).
\end{equation}
Based on Eqs.~\eqref{g2_diff} and~\eqref{g2_same}, the general expression for $g^{(2)}({\vb r}_1, {\vb r}_2)$ takes on the form
\begin{equation}
	g^{(2)}({\vb r}_1, {\vb r}_2) =  \left\{
	\begin{array}{ll}
		1, & \text{if } \nexists \Delta_l ({\vb r}_1, {\vb r}_2 \in \Delta_l),
		\\
		\frac{\langle n(n-1) \rangle_{Q}}{\langle n \rangle_{Q}^2}, & \text{if } \exists \Delta_l ({\vb r}_1, {\vb r}_2 \in \Delta_l).
	\end{array}
	\right.
\end{equation} 

In what follows, we will mostly focus on $g^{(2)}({\vb r}_1, {\vb r}_2)$ with ${\vb r}_1$ and ${\vb r}_2$ in the same cell.
First thing to note is that given ${\vb r}_1$ and ${\vb r}_2$ belong to the same cell, the two-particle density does not otherwise depend on the position of the particles, thus we can just write $\rho^{(2)}$ for simplicity. The same consequently applies to $g^{(2)}$.

Before proceeding with numerical and graphical results for the pair distribution function $g^{(2)}$ let us rewrite it in a few alternative representations.

If expressed in terms of $K_j$, quantities $\rho^{(2)}$ and $g^{(2)}$ take on the following forms
\begin{eqnarray}
	\rho^{(2)} = \frac{K_2(T^*, \mu^*; \bar{y}_{\rm max}) - K_1(T^*, \mu^*; \bar{y}_{\rm max})}{v^2 K_0(T^*, \mu^*; \bar{y}_{\rm max})},
\end{eqnarray}
and
\begin{eqnarray}
	g^{(2)} & = & \frac{\left[K_2(T^*, \mu^*; \bar{y}_{\rm max}) - K_1(T^*, \mu^*; \bar{y}_{\rm max})\right] K_0(T^*, \mu^*; \bar{y}_{\rm max})}{K_1(T^*, \mu^*; \bar{y}_{\rm max})^2},
\end{eqnarray}
respectively. See Appendix~\ref{sec:appendix_aux} for details.

We can also represent the results for the pair distribution function in the form of parametric equations. Applying the idea taken from~\cite{KD22}, we introduce a new function
\begin{equation}
	\label{eq:change_y_to_z}
	\bar{z}_{\rm max} = \bar{y}_{\rm max} + \frac{\mu^*}{T^*},
\end{equation}
and express the chemical potential as a function of two variables $(T^*, \bar{z}_{\rm max})$
\begin{equation}
	\label{eq:mu_in_z}
	\mu^* = T^* \bar{z}_{\mathrm{max}} - \frac{\tilde{K}_1(T^*;\bar{z}_{\mathrm{max}})}{\tilde{K}_0(T^*;\bar{z}_{\mathrm{max}})},
\end{equation}
where the special functions $\tilde{K}_j$ are written as
\begin{eqnarray}
	\tilde{K}_j(T^*;\bar{z}_{\mathrm{max}}) = \sum_{n=0}^{\infty} \frac{n^j (v^* T^{*3/2})^n}{n!} 
	\exp[\bar{z}_{\mathrm{max}}n - \frac{a}{2T^*}n^2].
\end{eqnarray}
Now, the two-particle density and the pair distribution function take on the following forms
\begin{equation}
	\rho^{(2)} = \frac{\tilde{K}_2(T^*; \bar{z}_{\rm max}) - \tilde{K}_1(T^*; \bar{z}_{\rm max})}{v^2 \tilde{K}_0(T^*; \bar{z}_{\rm max})},
\end{equation}
\begin{equation}
	\label{eq:g2_in_z}
	g^{(2)} = \frac{\left[\tilde{K}_2(T^*; \bar{z}_{\rm max}) - \tilde{K}_1(T^*; \bar{z}_{\rm max})\right] \tilde{K}_0(T^*; \bar{z}_{\rm max})}{\tilde{K}_1(T^*; \bar{z}_{\rm max})^2}.
\end{equation}
At any given temperature $T^*$, Eqs.~\eqref{eq:g2_in_z} and~\eqref{eq:mu_in_z} can be formally considered as a parametric equation for the pair distribution function $g^{(2)}$ as a function of $\mu^*$, with $\bar{z}_{\mathrm{max}}$ being the parameter.

Similarly, if Eq.~\eqref{eq:rho_in_y} is rewritten as
\begin{equation}
	\label{eq:rho_in_z}
	\rho^*(T^*;\bar{z}_{\mathrm{max}}) = \frac{\tilde{K}_1(T^*;\bar{z}_{\mathrm{max}})}{\tilde{K}_0(T^*;\bar{z}_{\mathrm{max}})},
\end{equation}
then Eqs.~\eqref{eq:g2_in_z} and~\eqref{eq:rho_in_z} constitute a parametric equation for the pair distribution function $g^{(2)}$ as a function of $\rho^*$, at any given $T^*$ and with $\bar{z}_{\mathrm{max}}$ being the parameter.

\begin{figure}[htbp]
	\includegraphics[width=0.5\textwidth,angle=0]{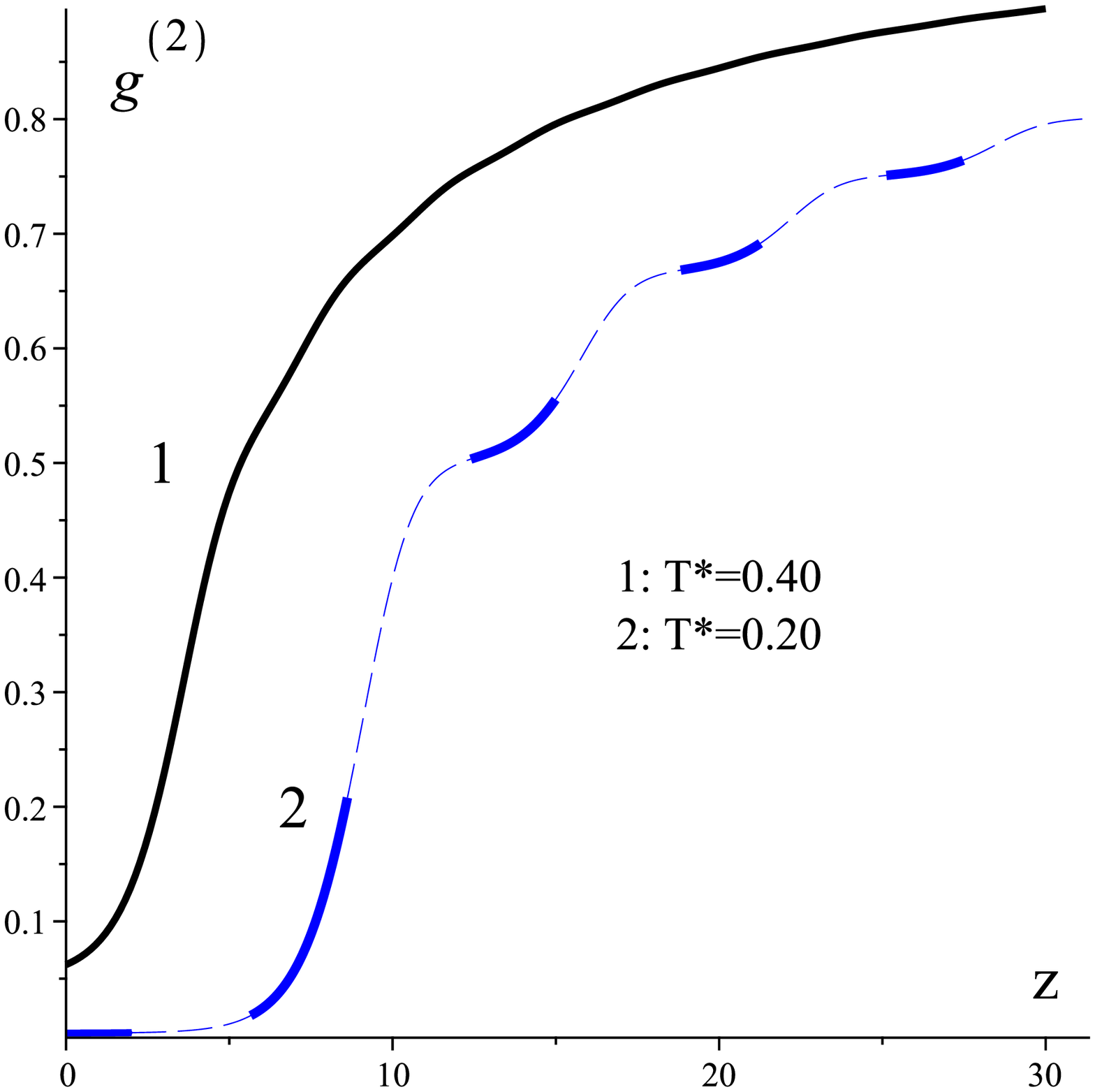}
	\hfill
	\includegraphics[width=0.5\textwidth,angle=0]{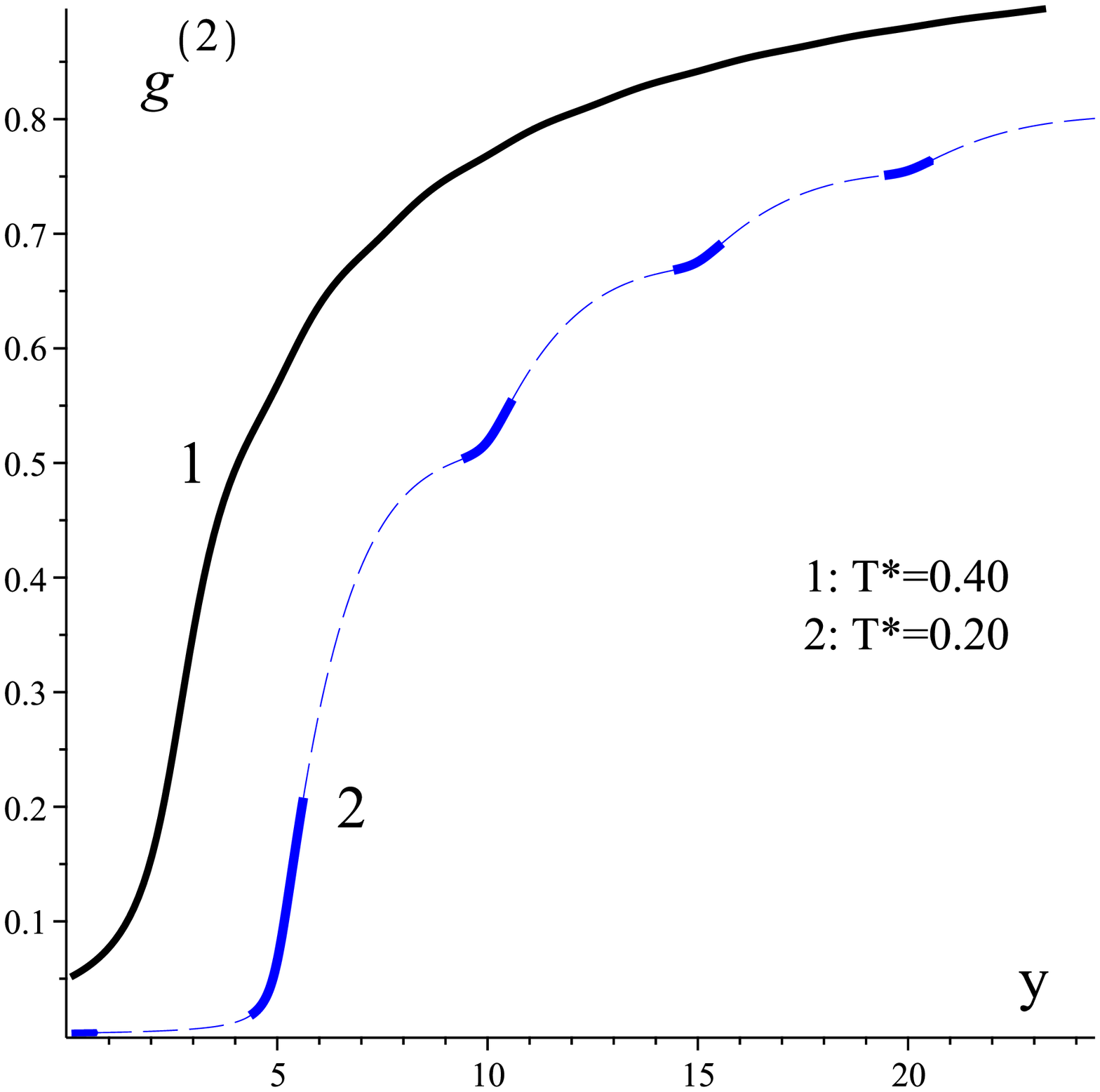}
	\\
	\parbox{0.45\textwidth}{\caption{\label{fig:g2_z} Pair distribution function $g^{(2)}$ versus $\bar{z}$ at two temperatures. Curve~1 (Black): $T^* = 0.4$, the temperature higher than $T^*_c$; Curve~2 (Blue): $T^*=0.2$, the temperature lower than $T^*_c$. The critical temperature $T^*_c \approx 0.25$. At Curve 2, the bold solid sections correspond to the stable phases, the dashed sections correspond to the metastable and unstable regions.}}
	\hfill
	\parbox{0.45\textwidth}{\caption{\label{fig:g2_y} Pair distribution function $g^{(2)}$ versus $\bar{y}$ at two temperatures. Curve~1 (Black): $T^* = 0.4$, the temperature is higher than $T^*_c$; Curve~2 (Blue): $T^*=0.2$, the temperature lower than $T^*_c$. The critical temperature $T^*_c \approx 0.25$. At Curve 2, the bold solid sections correspond to the stable phases, the dashed sections correspond to the metastable and unstable regions.}}
	
\end{figure}

\begin{figure}[htbp]
	\begin{subfigure}[b]{0.5\textwidth}
		\includegraphics[width=\textwidth,angle=0]{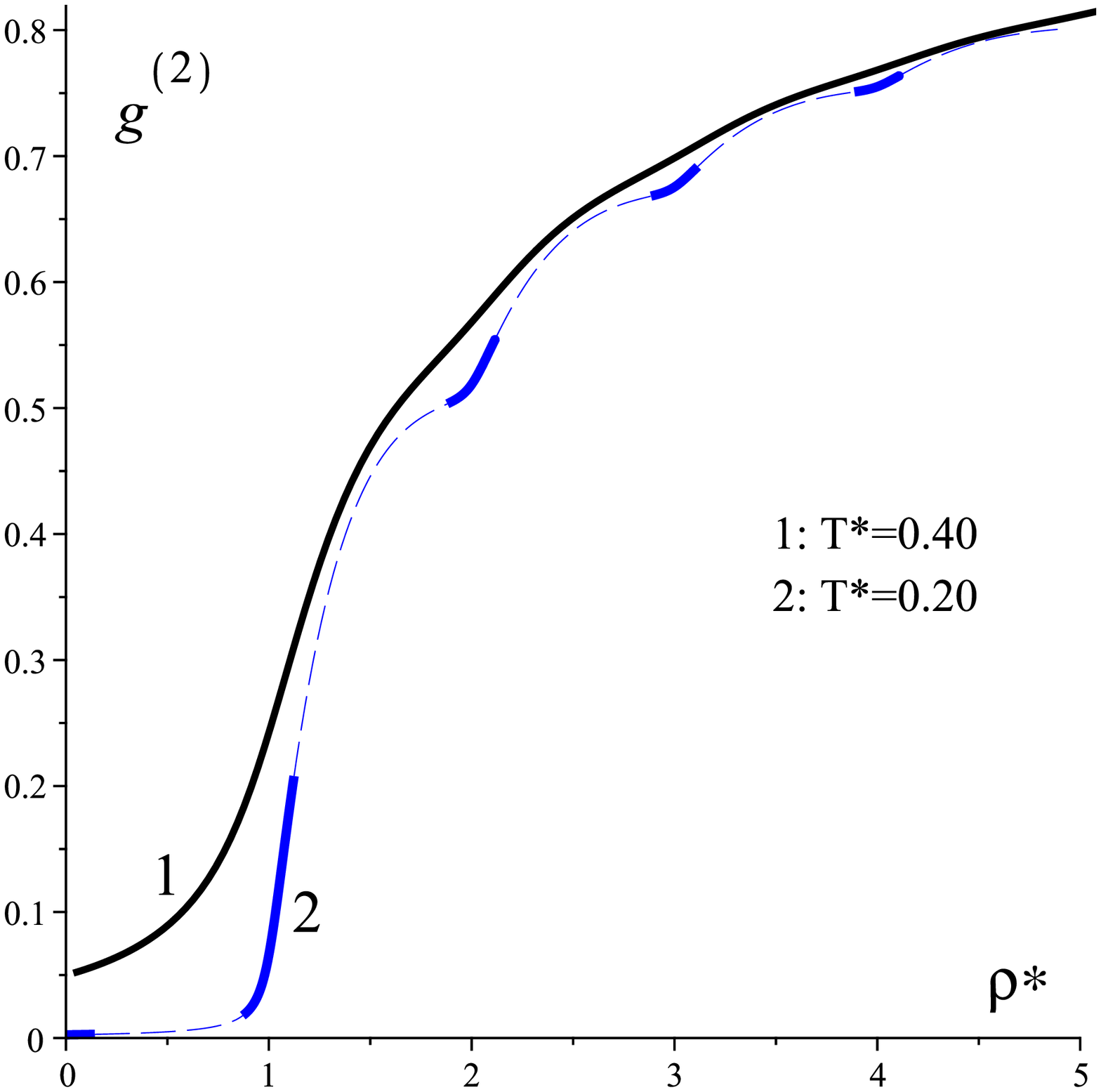}
		\caption{}
		\label{fig:g2_rho_a}
	\end{subfigure}
	\begin{subfigure}[b]{0.5\textwidth}
		\includegraphics[width=\textwidth,angle=0]{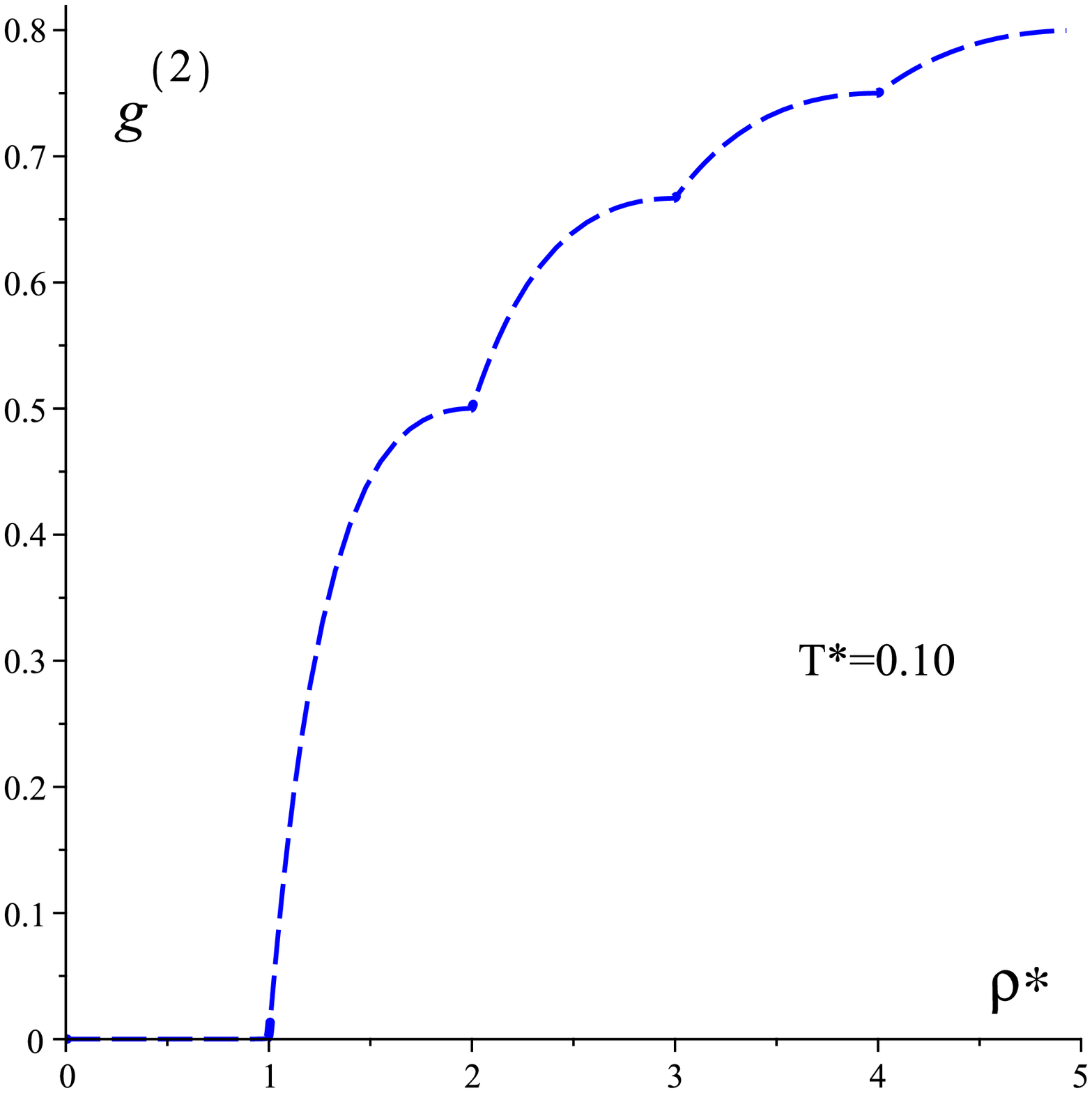}
		\caption{}
		\label{fig:g2_rho_low_temp_b}
	\end{subfigure}
	\caption{Pair distribution function $g^{(2)}$ versus $\rho^*$ at different temperatures.
	(\textbf{a}) Curve~1 (Black): $T^* = 0.4$ - temperature is higher than $T^*_c$; Curve~2 (Blue): $T^*=0.2$ - temperature is lower than $T^*_c$.
	(\textbf{b}) Blue curve: $T^*=0.1$ - temperature is much lower than $T^*_c$. The critical temperature $T^*_c \approx 0.25$. For undercritical temperatures, the bold solid sections correspond to the stable phases, the dashed sections correspond to the metastable and unstable regions. In Panel (b), the stable regions are too small to be noticeable in the plot.}
\end{figure}

At this point, we are in a position to illustrate the obtained results for the pair distribution function graphically. In Figure~\ref{fig:g2_z} the pair distribution function is shown as a function of $\bar{z}$ for two different values of temperature: $T^*=0.40$, which is above the range of critical temperatures, and $T^*=0.20$, which is below the range of critical temperatures. The values of critical temperatures are summarized in Table~\ref{tab1}. For the subcritical temperature $T^*=0.20$, the solid bold portions of the curve corresponds to the stable states, while the dashed ones correspond to the metastable or unstable states.

\begin{table*}
	\caption{Critical point coordinates for the first four critical points of the cell fluid model with Curie-Weiss interaction. The numerical calculations were performed for $a=1.2$ and $v^*=5.0$.}
	\label{tab1}
	\centering
	\begin{tabular}{ccccc}
		$n$ & $T^*_c$ & $\rho^*_c$ & $\mu^*_c$ & $\bar{z}_c$ \\
		\midrule
		
		1 & 0.254567 & 0.513896 & 0.209380 & 2.84119 \\
		\midrule
		2 & 0.261881 & 1.50557 & 0.585097 & 7.98327 \\
		
		\midrule
		3 & 0.265254 & 2.50303 & 0.891843 & 12.7985 \\
		
		\midrule
		4 & 0.267242 & 3.50191 & 1.16893 & 17.4779 \\
		
		\bottomrule
	\end{tabular}		
\end{table*}

In Figure~\ref{fig:g2_y} the pair distribution function is shown as a function of $\bar{y}$ for the same temperatures $T^*=0.4$ and $T^*=0.2$. The notation and legend is the same as in Figure~\ref{fig:g2_z}.

In Figure~\ref{fig:g2_rho_a} the pair distribution function is shown as a function of density $\rho^*$ for the same two temperatures, supercritical $T^*=0.40$ and subcritical $T^*=0.20$. We see that $g^{(2)}$ is a strictly increasing function of $\rho^*$ at a constant temperature. From the plot for the subcritical temperature, we can also conclude that the pair distribution function takes on higher values at each consecutive phase compared to the preceding one. Additionally, in Figure~\ref{fig:g2_rho_low_temp_b}, we display the pair distribution function for a very low temperature $T^*=0.1$. In such conditions, the stable phase intervals in $\rho^*$ are very narrow and almost invisible in the scale of the plot. Thus, the presented curve goes through different two-phase regions and corresponds to metastable and unstable states.

\begin{figure}[htbp]
	\centering
	\includegraphics[width=0.6\textwidth,angle=0]{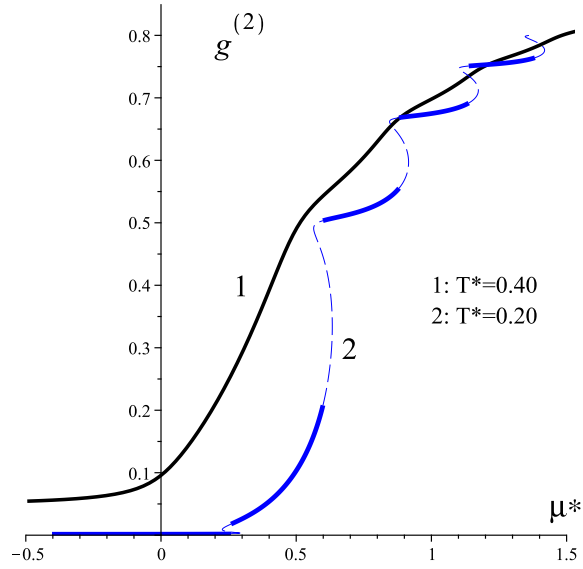}
	\caption{Pair distribution function $g^{(2)}$ versus $\mu^*$ at two different temperatures. Curve~1 (Black): $T^* = 0.4$, the temperature higher than $T^*_c$; Curve~2 (Blue): $T^*=0.2$, the temperature lower than $T^*_c$. The critical temperature $T^*_c \approx 0.25$. At Curve 2, the bold solid sections correspond to the stable phases, the dashed sections correspond to the metastable and unstable regions.}
	\label{fig:g2_mu}
\end{figure}

The dependence of the pair distribution function on the chemical potential $\mu^*$ is illustrated in Firugre~\ref{fig:g2_mu}. The behavior is exemplified for two temperatures $T^*=0.40$ and $T^*=0.20$.

\begin{figure}[htbp]
	\begin{subfigure}[b]{0.5\textwidth}
		\includegraphics[width=\textwidth,angle=0]{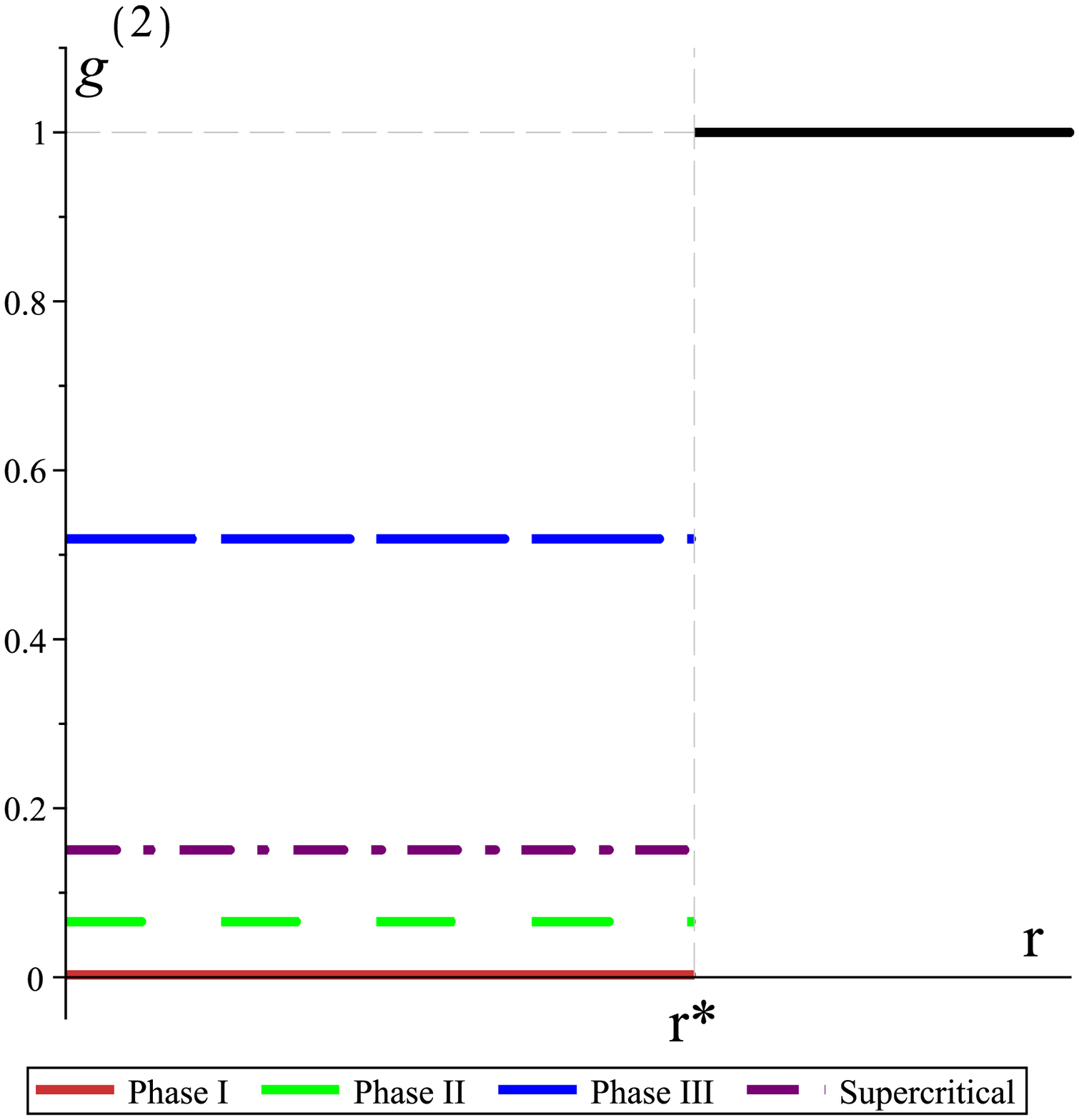}
		\caption{}
		\label{fig:g2_r_a}
	\end{subfigure}
	\begin{subfigure}[b]{0.5\textwidth}
		\includegraphics[width=\textwidth,angle=0]{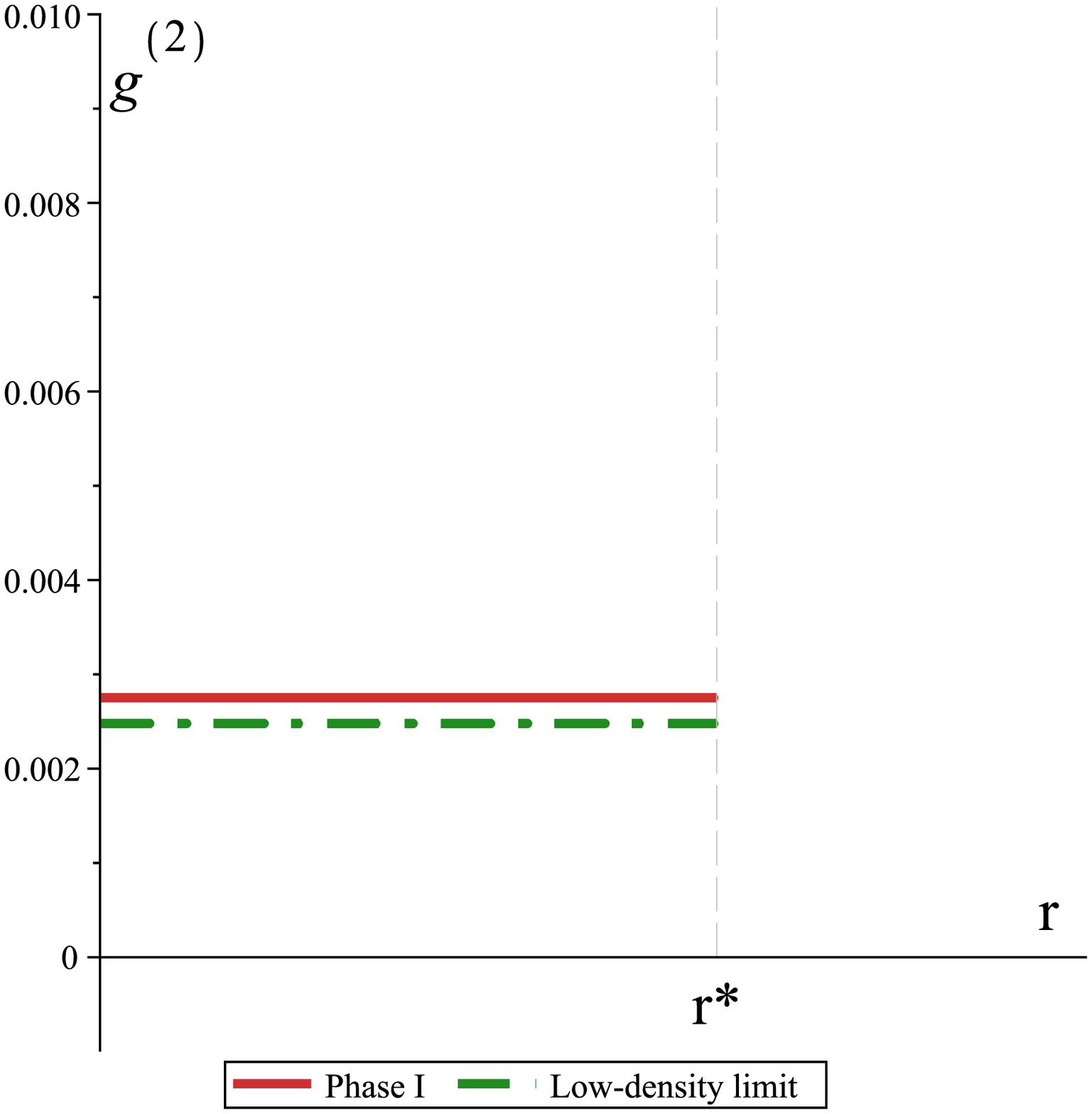}
		\caption{}
		\label{fig:g2_r_low_dens_b}
	\end{subfigure}
	\caption{Pair distribution function $g^{(2)}({\vb r}_1, {\vb r}_2)$ as a function of separation $r = \abs{{\vb r}_1 - {\vb r}_2}$. The quantity $r^*$ denotes distance from the point with coordinate ${\vb r}_1$ to the boundary of the cubic cell containing ${\vb r}_1$, in the direction of ${\vb r}_2$. ({\textbf{a}}) Phase I: $T^* = 0.2$, $\rho^*=0.1$; Phase II: $T^* = 0.2$, $\rho^* = 1.0$; Phase III: $T^*_c = 0.2$, $\rho^* = 2.0$; Supercritical: $T^* = 0.5$, $\rho^* = 0.5$. At $r > r^*$, $g^{(2)} = 1$. ({\textbf b}) Phase I: $T^* = 0.2$, $\rho^*=0.1$; Low-density limit: $g^{(2)} \approx \exp(-a/T^*)$, with $T^*=0.2$, $a=1.2$.}
\end{figure}

The spatial dependency of $g^{(2)}({\vb r}_1, {\vb r}_2)$ for the cell fluid model with Curie-Weiss interaction can be presented as a step function. While ${\vb r}_1$ and ${\vb r}_2$ are located within one cell, $g^{(2)}$ takes on a value that is dependent of the temperature and chemical potential (or of density). However as soon as the coordinates of two particles belong to different cells, $g^{(2)}$ becomes unity.
In Figure~\ref{fig:g2_r_a} such step dependency is illustrated for different values of temperature and density. The temperatures and densities are selected to correspond to Phase I, Phase II, Phase III, and to a supercritical region over the critical point for the Phase I -- Phase II coexistence. Since the value for Phase I is hardly visible in the scale of the plot, Figure~\ref{fig:g2_r_low_dens_b} shows $g^{(2)}$ in Phase I in a better scale, along with the low-density limit~\eqref{eq:g2_low_limit}, see Appendix~\ref{sec:low-dens} for details. For Phase I, $T^*=0.2$ and $\rho^*=0.1$ are selected as a representative state. For other states, the following values were selected: $T^*=0.2$ and $\rho^*=1.0$ for Phase II, $T^*=0.2$ and $\rho^*=2.0$ for Phase III, and, finally, $T^*=0.5$ and $\rho^*=0.5$ for the supercritical state.

Based on the obtained results, we see that the pair distribution function takes on its ideal-gas value as soon as the two particles are not located within the same cell. If two particle are indeed located within the same cell, then the value of $g^{(2)}$ becomes less than 1, which indicates that the probability of finding two particles at separation $r$ within a cell is smaller then would be in the case of an ideal gas. However, the pair distribution function does not otherwise depend on $r$, and therefore, no structure is observed even within a cell. Therefore, the cell fluid model with Curie-Weiss interaction is essentially a mean-field fluid.

One feature we should notice here is that the pair distribution function $g^{(2)}({\vb r}_1, {\vb r}_2)$ experiences a discontinuity at the cell boundary. This is similar to the behavior of the pair distribution function for the penetrable spheres model~\cite{LWL98}, where this quantity has a discontinuity at $r=\sigma$, $\sigma$ being the diameter of a sphere.

\section{Conclusions}
In this paper, we obtain exact expressions for the pair distribution function $g^{(2)}({\vb r}_1, {\vb r}_2)$ of the cell fluid model with Curie-Weiss interaction. Based on the obtained analytical result, dependences of $g^{(2)}$ on particle separation, density, and chemical potential are illustrated graphically. From the spacial dependency, we conclude that the pair distribution function takes on its ideal-gas value whenever the two particles occupy different cells. Only when both particles reside in the same cell does $g^{(2)}$ fall below unity, reflecting a reduced likelihood of finding two particles at separation $r$ compared with an ideal gas. Apart from this, the function exhibits no dependence on the interparticle separation $r$, and thus no intra-cell structure emerges. Consequently, the cell fluid with Curie-Weiss interaction is essentially a mean-field fluid.
From the density dependence, we see that the pair distribution function takes on higher values at each consecutive phase compared to the preceding one, which is expected as the phases are characterized by higher densities.

\bmhead{Acknowledgements} This work was supported by the National Research Foundation of Ukraine under the project No. 2023.03/0201.

\begin{appendices}

\renewcommand{\theequation}{A.\arabic{equation}}
\setcounter{equation}{0}

\section{Auxiliary calculations}
\label{sec:appendix_aux}

When ${\vb r}_1$ and ${\vb r}_2$ belong to the same cell, the two-particle density is expressed in terms of functions $K_j$ defined in~\eqref{def:Kj} as follows
\begin{eqnarray*}
	\rho^{(2)} & = & \frac{1}{v^2 K_0(T^*, \mu^*; \bar{y}_{\rm max})} \sum_{n=2}^{\infty} n(n-1) \frac{\left(v^* T^{*3/2}\right)^n}{n!} \exp\left[\left(\bar{y}_{\rm max} + \frac{\mu^*}{T^*}\right)n - \frac{a}{2T^*} n^2\right]
	\\
	& = & \frac{1}{v^2 K_0(T^*, \mu^*; \bar{y}_{\rm max})} \sum_{n=0}^{\infty} (n^2 - n) \frac{\left(v^* T^{*3/2}\right)^n}{n!} \exp\left[\left(\bar{y}_{\rm max} + \frac{\mu^*}{T^*}\right)n - \frac{a}{2T^*} n^2\right]
	\\
	& = & \frac{K_2(T^*, \mu^*; \bar{y}_{\rm max}) - K_1(T^*, \mu^*; \bar{y}_{\rm max})}{v^2 K_0(T^*, \mu^*; \bar{y}_{\rm max})}.
\end{eqnarray*}
Therefore, for $g^{(2)}$ one obtains
\begin{eqnarray}
	g^{(2)} & = & \frac{\left[K_2(T^*, \mu^*; \bar{y}_{\rm max}) - K_1(T^*, \mu^*; \bar{y}_{\rm max})\right] K_0(T^*, \mu^*; \bar{y}_{\rm max})}{K_1(T^*, \mu^*; \bar{y}_{\rm max})^2}
\end{eqnarray}

\section{\label{sec:low-dens}Low-density limit of the pair distribution function}
For an isotropic system the pair distribution function $g^{(2)}(\vb{r}_1, \vb{r}_2)$ is a function only of a separation $r \equiv \abs{\vb{r}_1 - \vb{r}_2}$, and the low-density limit is equal to the Boltzmann factor $e(r)$ of a pair interaction potential $v(r)$, see e.g. Section~2.6 in~\cite{HansenMcDonald13}:
\begin{equation*}
	\lim_{\rho \to 0} g^{(2)}(r) = e(r) \equiv \exp(-\beta v(r)).
\end{equation*}
Let us assume that in the case of the cell model studied in this paper, the following comes true as well
\begin{equation}
	\lim_{\rho \to 0} g^{(2)}({\vb r}_1, {\vb r}_2) = e({\vb r}_1, {\vb r}_2)
\end{equation}
where the Boltzmann factor for the Curie-Weiss interaction defined in~\eqref{def:curie-weiss-pot} is
\begin{equation*}
	e({\vb r}_1, {\vb r}_2) = \exp\left[-\beta \Phi_{N_v}({\vb r}_1, {\vb r}_2)\right].
\end{equation*}
If ${\vb r}_1$ and ${\vb r}_2$ are in different cells, then
\begin{equation*}
	e({\vb r}_1, {\vb r}_2) = \exp(\frac{\beta J_1}{N_v}) = \exp(\frac{\beta^*}{N_v}).
\end{equation*}
where $\beta^* = 1/T^*$.
In the limit of $N_v \to \infty$
\begin{equation*}
	e({\vb r}_1, {\vb r}_2) \approx 1 + \frac{\beta^*}{N_v} \approx 1.
\end{equation*}
If ${\vb r}_1$ and ${\vb r}_2$ are in the same cell, then
\begin{equation*}
	e({\vb r}_1, {\vb r}_2) = \exp(\frac{\beta J_1}{N_v} - \beta J_2) = \exp(\frac{\beta^*}{N_v} - \beta^* a). 
\end{equation*}
In the limit of $N_v \to \infty$,
\begin{equation*}
	e({\vb r}_1, {\vb r}_2) \approx \exp(-\beta^* a).
\end{equation*}
Therefore, 
\begin{equation*}
	\lim_{\rho \to 0} g^{(2)}({\vb r}_1, {\vb r}_2) = \left\{
	\begin{array}{ll}
		\exp[\frac{\beta^*}{N_v} - \beta^* a], \quad & \text{if } \exists \Delta_l ({\vb r}_1, {\vb r}_2 \in \Delta_l),
		\\
		\exp[\frac{\beta^*}{N_v}], \quad & \text{if } \nexists \Delta_l ({\vb r}_1, {\vb r}_2 \in \Delta_l).
	\end{array}
	\right.
\end{equation*}
which in the thermodynamic limit $N_v \to \infty$ becomes
\begin{equation}
	\label{eq:g2_low_limit}
	\lim_{\substack{\rho \to 0 \\ N_v \to \infty}} g^{(2)}({\vb r}_1, {\vb r}_2) = \left\{
	\begin{array}{ll}
		\exp[- \beta^* a], \quad & \text{if } \exists \Delta_l ({\vb r}_1, {\vb r}_2 \in \Delta_l),
		\\
		1, \quad & \text{if } \nexists \Delta_l ({\vb r}_1, {\vb r}_2 \in \Delta_l).
	\end{array}
	\right.
\end{equation}




\end{appendices}





\end{document}